\documentclass[12pt]{iopart}

\usepackage{iopams}
\expandafter\let\csname equation*\endcsname\relax
\expandafter\let\csname endequation*\endcsname\relax
\usepackage{amsmath}
\usepackage{tensor}
\usepackage[colorlinks=true,linkcolor=blue,citecolor=blue,urlcolor=blue]{hyperref}
\newcommand{\eqr}{\eqref}

\usepackage{graphicx}     
\usepackage[dvipsnames]{xcolor}
\usepackage{floatflt}     
\usepackage{wrapfig}
\usepackage{color}
\usepackage{pstricks}
\usepackage{subcaption}
\usepackage{color}
\definecolor{DarkRed}{rgb}{0.35,0.01,0.01}
\definecolor{Linen}{rgb}{0.98,0.98,0.94}
\definecolor{Blue}{rgb}{0.,0.,1.0}
\definecolor{DarkBlue}{rgb}{0.099,0.099,0.44}
\definecolor{DarkGreen}{rgb}{0.0,0.4,0.0}
\definecolor{Turquoise}{rgb}{0.0,0.9,0.7}
\usepackage{tikz}
\usepackage{tkz-euclide}
\usepackage{tikz-3dplot}
\usepackage{pgfplots,pgfplotstable}
\usepackage[makeroom]{cancel}
\pgfplotsset{compat=newest}
\usetikzlibrary{shapes,shapes.arrows,shapes.symbols,shapes.geometric, calc,positioning}
\usepackage{mathtools}
\tikzstyle{start} = [rectangle, rounded corners, 
minimum width=3cm, 
minimum height=0.7cm,
text centered, 
text width=3cm, 
draw=black, 
fill=orange!30]

\tikzstyle{goal2} = [rectangle, rounded corners, 
minimum width=3cm, 
minimum height=0.7cm,
text centered, 
text width=3cm, 
draw=black, 
fill=red!30]

\tikzstyle{goal1} = [rectangle, rounded corners, 
minimum width=3cm, 
minimum height=0.7cm,
text centered, 
text width=3cm, 
draw=black, 
fill=purple!30]

\tikzstyle{goal3} = [rectangle, rounded corners, 
minimum width=3cm, 
minimum height=0.7cm,
text centered, 
text width=2.5cm, 
draw=black, 
fill=green!30]
\tikzstyle{arrow} = [thick,->,>=stealth] 
\begin{document}

\title[Geometrodyn. of a 2D C. Surface due to a Constr. Quan. Particle]{Geometrodynamics of a 2D Curved Surface due to a Constrained Quantum Particle via its Gravitational Dual{: $\mathbf{\mathcal{S}^2}$ Analytical Model Calculations}}

\author{Shanshan Rodriguez$^{1,2,*}$, Leo Rodriguez$^{1,2}$, Zhenzhong Xing$^3$, Connor McMillin$^{1}$, L. R. Ram-Mohan$^{2}$}

\address{$^1$ Department of Physics, Grinnell College, Grinnell IA, 50112, USA}
\address{$^2$ Department of Physics, Worcester Polytechnic Institute, Worcester MA, 01609, USA}
\address{$^3$ Department of Electrical and Computer Engineering, Cornell University, Ithaca, NY 14850}
\address{$^*$ Author to whom any correspondence should be addressed.}
\ead{rodriguezs@grinnell.edu}
\vspace{10pt}
\begin{indented}
\item[]\today
\end{indented}

\begin{abstract}
We provide a unique and novel extension of da Costa's calculation of a quantum mechanically constrained particle. 
This is achieved by analyzing the perturbative back reaction of the quantum confined particle's eigenstates and spectra upon the geometry of the curved surface itself, thereby addressing the problem of shape optimization in this model. We do this by first formulating a two-dimensional action principle of the quantum constrained particle, which upon {variation of the wave function} reproduces Schr\"odinger's equation including da Costa's surface curvature-induced potentials. We further demonstrate that our derived action principle is dual to a two-dimensional dilation gravity theory and we vary its functional with respect to the embedded two-dimensional inverse-metric to obtain the respective geometrodynamical Einstein equation. We solve this resulting Einstein equation perturbatively by first solving the da Costa's Schr\"odinger equation to obtain an initial eigensystem, which is used as initial-input data for a perturbed metric inserted into the derived Einstein equation. As a proof of concept, we perform this calculation on a two-sphere and show its first iterative perturbed shape evolution. {We also turn on external electromagnetic fields and formulate the full field theoretic field equations for future investigation. The external fields manifest themselves via a surface induced, pulled-back $U(1)$ coupling in our two-dimensional dual gravity theory, thereby revealing interesting and rich new surface physics in this specific paradigm.}
\end{abstract}
%
\vspace{1pc}
\\\submitto{\PS}
%
%
%
\section{Introduction}
{Since da Costa's seminal work and Jensen and Koppe's independent calculation} \cite{PhysRevA.23.1982,JENSEN1971586} demonstrating the emergence of curvature-induced surface potentials within Schr\"odinger's equation, the importance of differential geometric considerations in quantum semiconductor systems has spawned a plethora of exciting research areas. These curvature-induced surface potentials arise while expanding and separating the Laplace-Beltrami operator over $\mathbb{R}^3$ and confining a quantum particle to an embedded curved two-dimensional surface $\mathcal{M}^2$. Taking $\mathcal{M}^2$ to the nanoscale opens a wide area of curvature-induced quantum phenomena ranging from curved-spintronics, geometric nonlinear Hall effects, curvilinear magnetism, to strain-induced geometric potentials, to name a few \cite{electricMatRev}. {Additionally, the curvature of the nano surfaces \cite{PhysRevB.81.205409}, such as graphene nanotori and Buckyballs \cite{GeoEffCurvNano,EleGraNan}, diffusive ferromagnetic nanowires \cite{PhysRevB.105.134511}, and other interesting geometries \cite{MA2022107621,PhysRevD.90.025006} etc., feeds back onto the quantum electronic material properties leading to an optimization and tuning problem}. 

{Now, since the da Costa paradigm (which generalizes Jensen and Koppe's work to arbitrary $\mathcal{M}^2$) necessitates differential geometric considerations \cite{Kreyszig} in quantum systems \cite{Kleinert:1997em}, {it sets the stage for the implementation of exotic wormhole \cite{CasWormHole,PhysRevB.106.165426}, two-dimensional black holes \cite{SCHMIDT2023169465,COSTAFILHO2021114639} and other gravity-inspired two-dimensional curved spaces \cite{ATANASOV2020126042,Biswas:2019owq} relevant to two-dimensional material semiconductor/condensed matter applications \cite{PhysRevD.108.086006,doi:10.1142/S0219887820501352,PhysRevA.107.062806}. {The above hints towards a novel string-theoretic pathway of addressing the important aforementioned question about geometric shape optimization for curved two-dimensional semiconductor devices} \cite{electricMatRev}}. This is the premise of our work in this {article}}. 

{The important question of shape optimization and perturbative shape evolution of semiconductor devices due to external fields and/or quantum mechanical processes on their surfaces has been of great interest in the above-mentioned and many other related research fields \cite{ram3d,cite-key,YAMADA20102876}}. In particular, our collaboration {has addressed shape optimization} from a fundamental field theoretic approach in another related condensed matter picture, namely that of the extraordinary magnetoresistance effect in hybrid semiconductor-metal structures, which is highly dependent upon semiconductor device geometry {\cite{Rodriguez:2019djz}}. In this specific case, the geometrodynamic features reveal themselves in the resistivity tensor $\tilde{\mathcal{G}}_{\mu\nu}$, which may be decomposed into three fields:
\begin{align}
\tilde{\mathcal{G}}_{\mu\nu}=e^{2\psi_0}\left(g_{\mu\nu}+\beta_{\mu\nu}\right),
\end{align}
where $\psi_0$ relates to the conductivity at zero external magnetic field, $g_{\mu\nu}$ is the metric tensor and describes the geometry of the respective semiconductor devise, and $\beta_{\mu\nu}$ is the external magnetic field. A visual analog inspection of $\tilde{\mathcal{G}}_{\mu\nu}$ above immediately reminds us of a conformally scaled generalized geometric metric {with bosonic string theory field content. In this analogy, $\psi_0$ would be some fixed value of the dilation $\psi$ and $\beta_{\mu\nu}$ would be the Kalb-Ramond field with critical (conformal-anomaly free)} closed-string effective action, after renormalization group flow analysis, given by:
\begin{align}
S_{\rm{eff}}=-\frac{1}{2\kappa}\int d^{26} x\sqrt{-g}e^{2\psi}\left\{\mathcal{R}-4(\nabla\psi)^2+\frac{1}{12}H^2\right\},
\end{align}
{where $H=d\beta$, called the contortion or Ramond-Neveu-Schwarz three-form, and $H^2=H^\star\wedge H$}. However, since extraordinary magnetoresistance is not motivated from string theory, a more fundamental approach rooted in generalized geometry {\cite{Vysoky:2015psz}} is appropriate. This approach naturally encodes string theory's $B$-field symmetry and allows us to construct an action principle in which both geometry, $g_{\mu\nu}$, and external magnetic field, $\beta_{\mu\nu}$, are dynamical variables/fields and thus addresses the shape optimization of the extraordinary magnetoresistance phenomenon.

{Within this generalized geometric framework the generalized curvature may be obtained in terms of the Levi-Civita one and the contortion:}
\begin{align}
R_{\mu\nu}=R^{LC}_{\mu\nu}-\frac12\nabla^{LC}_\alpha H\indices{_\mu_\nu^\alpha}-\frac14H\indices{_\nu_\beta^\alpha}H\indices{_\alpha_\mu^\beta},
\end{align}
from which the conformally generalized geometric action of the extraordinary magnetoresistance may be constructed, yielding {\cite{Rodriguez:2019djz}}: 
\begin{align}
S_{\tilde{\mathcal{G}}}=\frac{e^{2\psi_0}}{2\kappa^2}\int d^4 x\sqrt{-g}\left\{\mathcal{R}^{LC}-\frac{1}{4}H^2\right\}.
\end{align}
The above can be interpreted as a low energy effective four-dimensional string theory and allows for a dynamical metric and the computation of the respective Einstein equation, via metric variation, which addresses shape optimization. The action can then be further reduced in order to model the extraordinary magnetoresistance with optimized geometry in lower dimensional semiconductors {\cite{Rodriguez:2019djz}}. {The popularity and utility of string theory applications in condensed matter is well established, most within the $AdS/CMT$\footnote{Anti de Sitter space/Condensed matter theory, a subfield of the overarching $AdS/$Conformal field theory correspondence of string theory.} correspondence {\cite{Sachdev:2010ch,Maldacena:1997re}}. We will show that da Costa's paradigm, when addressing the shape optimization mentioned above, will also benefit greatly from a string duality/correspondence. }

In contrast to the above discussed extraordinary magnetoresistance phenomenon and the general $AdS/CMT$ correspondence, the da Costa case finds its stringy connection in the fact that a quantum particle is confined to a two-dimensional curved surface. Then, when asking about shape optimization of the two-dimensional curved geometry, string theory arises naturally. This is due to the case that in two dimensions the curvature of any Riemannian Levi-Cevita connection two-form $\omega_{\alpha\beta}$ is given by; $d\omega_{12}=K vol^2$ and is thus completely specified by its Gauss curvature, $K=\frac{1}{2(2-1)}\mathcal{R}$, where $\mathcal{R}$ is the Ricci scalar. {An additional consequence of this fact is that the Einstein tensor must be identically zero in two dimensions.} In other words, there are no classical general relativistic dynamics in two dimensions, which is apparent since the Gauss-Bonnet theorem states that the Einstein-Hilbert action in two dimensions is proportional to the topological invariant Euler character, assuming the geometries in question are {handlebodies} in two dimensions. Thus, a two-dimensional shape optimization problem in semiconductor or CMT applications requires some type of quantum gravitational analogue, since there are no classical general relativistic geometrodynamics in the respective dimension. Such two-dimensional analogues arise naturally in the near horizon regime of black holes, which manifest themselves as two-dimensional world sheet and one dimensional bulk string theories and are commonly referred to as \emph{two-dimensional dilation gravities} {\cite{strom1}}. 

The specific proposed premise of this paper is to consider confining a quantum mechanical particle to the two-dimensional surface of a curved semiconductor or nano surface (such as a Buckyball/carbon-nanotube, etc.) and analyze how the particles' resulting eigensystem feeds back onto the geometry of the curved surface. In other words, is there an optimized nano geometry for a given quantum particle to be confined to, which would represent the total system's lowest eigenenergy configuration and therefore minimize any externally required work for confinement. Knowledge of such optimized geometries could have significant application in future device design and manufacturing. 

Our preliminary results demonstrate that there is a quantum confined eigensystem feedback which generates geometric deformations (a geometric strain of some kind) of the originally chosen two-dimensional semiconductor surface. It thus becomes an important task to compute these deformations in order to discover, or perturbatively narrow down, the ideal optimized geometric shape for the two-dimensional curved surface. Additionally, in the presence of external fields, such as electromagnetic fields, the geometric deformations on the semiconductor attain additional {nuance} with rich physics. The full geometrodynamics of this rich physical scenario turns out to be fully captured by a dual two-dimensional dilaton gravity, nearly identical to the two-dimensional quantum gravities that arise in the near horizon of four dimensional classical black holes. 

Our starting point is {the work} of da Costa \cite{PhysRevA.23.1982}, who showed that for a particle quantum mechanically confined to a two-dimensional curved surface embedded in $\mathbb{R}^3$, the pullback of Schr\"odinger's equation results in the appearance of surface potentials that depend on the two-dimensional curved surface's Ricci scalar and also the extrinsic curvature. Thus, motivated by the high accuracy action-level finite element method (FEM) {\cite{Ram}}, we formulate a two-dimensional action principle of the quantum constrained particle. Upon wave function variation, this formulation reproduces Schr\"odinger's equation according to da Costa, providing us with a foundational starting point to a full geometrodynamical description of this physical picture. 

\begin{figure}[h!] 
	\centering
	\includegraphics[width=.75\textwidth]{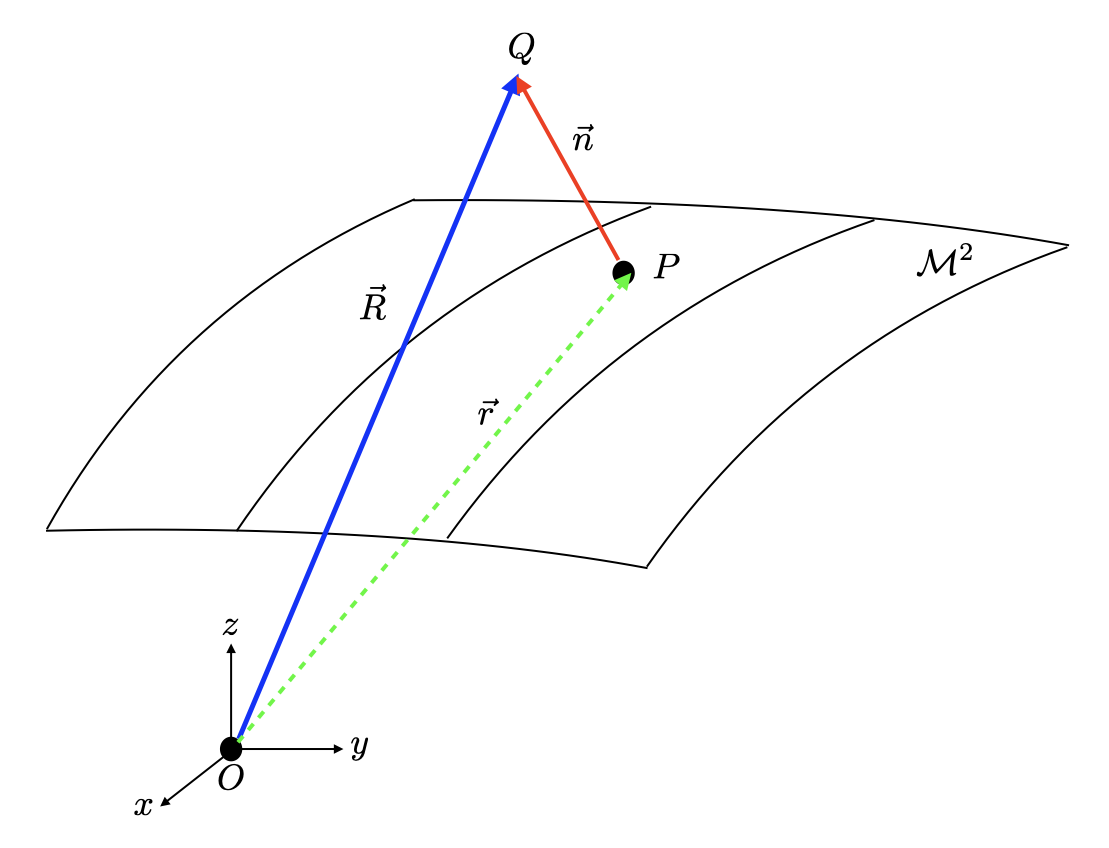}
\caption{\label{fig:M2EmR3}{Depiction of the coordinate map from $O$ to $Q$ via the immediate neighborhood of points $P\in\mathcal{M}^2$.}}
  \end{figure}

{Next, we will quickly review da Costa's {main} result in \cite{PhysRevA.23.1982} following the detailed and more modern pedagogical presentation provided in {\cite{Mazharimousavi:2021fyj}}.} Let $\vec R$ be the position vector of any point $Q$ $\in$ $\mathbb{R}^3$, such that $\vec R=\vec R(x,y,z)$. {Also, let $\vec r=\vec r(\theta,\phi)$ be the parametric representation of points contained in an immediate neighborhood of a curved two-dimensional surface $\mathcal{M}^2$, embedded in $\mathbb{R}^3$}. To reach $Q$ from $\mathcal{M}^2$ we choose a perpendicular distance $\rho$ from $\mathcal{M}^2$ to $Q$. Thus $\vec R=\vec r+\rho\hat n$, where $\hat n$ is the unit normal to $\mathcal{M}^2$ depicted in Fig.~\ref{fig:M2EmR3}. 
We will define the three dimensional Euclidean metric in the usual way:
\begin{align}
g_{\mu\nu}=\left(\partial_\mu\vec R\right)\cdot\left(\partial_\nu\vec R\right),
\end{align}
and the metric on $\mathcal{M}^2$ (first fundamental form) given by:
\begin{align}
h_{ab}=\left(\partial_a\vec r\right)\cdot\left(\partial_b\vec r\right). 
\end{align}
Here, greek indices range over the three Cartesian coordinates and latin indices range over the two embedded curved coordinates. We also define the extrinsic curvature (second fundamental form) given by:
\begin{align}
k_{ab}=\left(\partial_a\vec r\right)\cdot\left(\partial_b\hat n\right),
\end{align}
which depends on how $\vec r$ is embedded in $\mathbb{R}^3$ and informs about the rate of change of $\hat n$ as it is parallel transported across $\mathcal{M}^2$.

Now, given the above definitions we have:
\begin{align}
g_{ab}=\left(\partial_a\vec r+\rho\partial_a\vec n\right)\cdot\left(\partial_b\vec r+\rho\partial_b\vec n\right),
\end{align}
and employing the Weingarten equation {\cite{Kreyszig,both}}:
\begin{align}
\partial_a\vec n=k\indices{^b_a}\partial_b\vec r,
\end{align}
it can be shown that:
\begin{align}
g_{ab}=h_{ab}+2\rho k_{ab}+\rho^2 k\indices{^c_a}k\indices{^d_b}h_{cd},
\end{align}
and furthermore that: 
\begin{align}
g_{\mu\nu}=\left(\begin{array}{cc}g_{ab} & 0 \\0 & 1\end{array}\right).
\end{align}
Next, combining the two above results and computing the determinant $g=\det\left( g_{\mu\nu}\right)$ we obtain \cite{PhysRevA.23.1982}:
\begin{align}
g=h\left(1+\rho \mathcal{K}+\rho^2 \mathcal{R}/2\right)^2=h\gamma^2,
\end{align}
where $h$ is the determinant of $h_{ab}$, $\mathcal{K}$ is the trace of the extrinsic curvature ({principal} curvature) and $\mathcal{R}$ is the Ricci scalar curvature (twice the Gaussian curvature) of $\mathcal{M}^2$. Next, considering this result and the three dimensional steady state Schr\"odinger equation, we have:
\begin{gather}
-\frac{\hbar^2}{2m}\frac{1}{\sqrt{g}}\partial_\mu\left(\sqrt{g}g^{\mu\nu}\partial_\nu\psi\right)+V_0\tilde{\delta}(\rho)\psi=E\psi\\\label{Eq:OrigSchEQ}
-\frac{\hbar^2}{2m}\frac{1}{\sqrt{h}\gamma}\partial_a\left(\sqrt{h}\gamma g^{ab}\partial_b\psi\right)-\frac{\hbar^2}{2m}\frac{1}{\gamma}\partial_\rho\left(\gamma \partial_\rho\psi\right)+V_0\tilde{\delta}(\rho)\psi=E\psi
\end{gather}
{where $\tilde{\delta}(r)$ is chosen such that $V(\rho)=V_0\tilde{\delta}(\rho)$ confines the particle to the two-dimensional surface. A physically interesting example, in addition to da Costa's original one, is provided in \cite{Mazharimousavi:2021fyj} and given by:}
\begin{align}
\tilde{\delta}(\rho)=
\begin{cases}
0&0<\rho<\epsilon\\
\infty&elsewhere
\end{cases}.
\end{align}
{In the above, $\epsilon$ represents the thickness of the nano surface, which presents interesting additional exploratory avenues of the radial equation including a closed form energy eigenspectrum \cite{Mazharimousavi:2021fyj} and a possible evanescent wave mode analysis {\cite{10.1063/1.5084052}}.}
Finally, we are able to separate the three dimensional equation \eqref{Eq:OrigSchEQ} into two equations via the ansatz $\psi=\frac{1}{\sqrt{\gamma}}\psi_{2D}(\theta,\phi)R(\rho)$, yielding \cite{PhysRevA.23.1982}:
\begin{gather}
-\frac{\hbar^2}{2m}\frac{1}{\sqrt{h}}\partial_a\left(\sqrt{h}h^{ab}\partial_b\psi_{2D}\right)-\frac{\hbar^2}{2m}\left\{\left(\mathcal{K}/2\right)^2-\mathcal{R}/2\right\}\psi_{2D}=E_{2D}\psi_{2D}\label{eq:DC2d}\\
-\frac{\hbar^2}{2m}R''+V_0\tilde{\delta}(\rho)R=E_\rho R,
\end{gather}
where the ansatz is chosen; given the normalization condition such that: 
\begin{align}
\int |\psi|^2dV=\int \left|\frac{1}{\sqrt{\gamma}}\psi_{2D}(\theta,\phi)R(\rho)\right|^2\sqrt{h}\gamma d^3x=1\Rightarrow
\begin{cases}
\int|\psi_{2D}|^2\sqrt{h}d^2x&=1\\
\int|R|^2d\rho&=1
\end{cases}.
\end{align}
\section{Two-Dimensional Gravity Duality}
{Our first task is to construct an effective quantum action, following techniques outlined in {\cite{PhysRev.82.914,RevModPhys.29.377,Ram}}, whose equation of motion is given by \eqref{eq:DC2d} after functional variation with respect to $\psi_{2D}$. In this scenario $\psi_{2D}$ may be interpreted as an auxiliary field introduced to restore locality within the path integral formulation of quantum mechanics, i.e.:
	\begin{align}
	\begin{split}
	\psi(x,t)=&\int dx_0\langle x|e^{-iHt/\hbar}|x_0\rangle\langle x_0|\psi\rangle\\
	=&\int_{x_0}^x \mathcal{D}\chi e^{-\frac{i}{\hbar}S_{cl}[\chi]}\langle x_0|\psi\rangle\\
	=&\int_{x_0}^x \mathcal{D}\chi e^{-\frac{i}{\hbar}S_{cl}[\chi]}\psi(x_0).
	\end{split}
	\end{align}
In the above, the propagator $\langle x|e^{-iHt/\hbar}|x_0\rangle$ has been replaced by the infinite dimensional integral over all possible paths $\chi$, of the Boltzmann factor containing the classical action $S_{cl}[\chi]$ {\cite{RevModPhys.29.377,larkQM}}, both of which are non-local quantities. To illustrate the quantum effective action interpretation we compute the above convolution on the right for $t\in\{0,T\}$, implementing the expansion 
	\begin{align}
	\begin{split}
	\lim_{T\to0}e^{\frac{im}{2\hbar}\frac{\left(x-x_0\right)^2}{T}}=&\sum_{i=0}c_i\left(\frac{d}{dx}\right)^i\delta(x-x_0)
	\end{split}
	\end{align}
for the yet-to-be determined constants $c_i$, and expanding the left to order $T$ within the same limit, yielding:
	\begin{align}
	\begin{split}
	\psi(x)+T\dot\psi(x,t)=&\psi(x)-\frac{\hbar}{i2m}T\psi''(x)-\frac{iT}{\hbar}V(x)\psi(x)+\mathcal{O}\left(T^2\right).
	\end{split}
	\end{align}
The above implies, for general time, that $\psi$ solves the Schrödinger equation: 
	\begin{align}
	\begin{split}
	i\hbar\dot\psi=&-\frac{\hbar^2}{2m}\psi''+V(x)\psi,
	\end{split}
	\end{align}
which demonstrates the equivalence between the path integral formulation of quantum mechanics, $Z=\int_{x_0}^x \mathcal{D}\chi e^{-\frac{i}{\hbar}S_{cl}[\chi]}$ and the above Schrödinger equation. In other words, within the framework of effective quantum theory, the above equivalence implies {\cite{PhysRev.82.914,RevModPhys.29.377,Button:2010kg}}: 
	\begin{align}
	\begin{split}
	\int\mathcal{D}\chi e^{-\frac{i}{\hbar}S_{cl}[\chi]}\sim&e^{iS_{\rm{eff}}(\psi)},
	\end{split}
	\end{align}
where $S_{\rm{eff}}(\psi)$ is the quantum effective action of the auxiliary field $\psi$. Now since $\psi$ solves the Schrödinger equation, $S_{\rm{eff}}(\psi)$ must be such that $\delta_\psi S_{\rm{eff}}(\psi)=0$ yields the respective Schrödinger equation as the quantum effective Euler-Lagrange equation of motion.} 

Assuming without loss of generality, that $\psi_{2D}$ is real {(which we will comment on later)} as this does not affect the final version of Schr\"odinger's equation in \eqref{eq:DC2d}, {we obtain:
\begin{align}
\begin{split}
S_{\rm{eff}}=&\frac{\hbar^2}{4m}\int d^2x\sqrt{h}\left\{\partial_a\psi_{2D}\partial^a\psi_{2D}+\left[\mathcal{R}/2-\left(\mathcal{K}/2\right)^2\right]\psi_{2D}^2-\epsilon\psi_{2D}^2\right\}\label{eq:2DSchAct},
\end{split}
\end{align}}
where $\epsilon$ is the scaled dimensionless energy $E_{2D}$. The above action {would now constitute a viable starting point for employing a finite element analysis in order to solve the Schr\"odinger equation \eqref{eq:DC2d}, thereby obtaining eigenstates and eigenvalues} for a number of different curved two-dimensional geometries. {Additionally, the above action can also be utilized to explore shape optimization (geometrodynamical evolution) of the two-dimensional embedded semiconductor surface, due to the back reaction of the quantum eigensystem. This can be achieved by elevating the embedded metric, $h_{ab}$, to a dynamical field variable and performing a functional variation with respect to its inverse, $h^{ab}$ (for canonical reasons), yielding an Euler-Lagrange equation of motion for the embedded metric. Such field equations resulting from inverse-metric variations are commonly referred to as Einstein equations, whose solutions represent optimized geometries \cite{strom1}. The computation of the respective novel Einstein equation for the da Costa paradigm is one of our main calculational results, unique to our work and not considered in past literature.} 

{However, before performing the above variation, it is very useful to rescale the wave function in terms of an analog dilaton field $\varphi$, given by $\psi=e^{-\varphi}$. This rescaling
reveals another one of our unique and novel results, given by the new action below and resulting da Costa/gravity duality (also previously unexplored in the relevant literature). Additionally, the resulting duality significantly improves the very lengthy calculation of the respective Einstein equation. Given this rescaling, the effective action \eqref{eq:2DSchAct} takes the form:
\begin{align}
S_{\rm{eff}}=\frac{\hbar^2}{4m}\int d^2x\sqrt{h}e^{-2\varphi}\left\{\partial_a\varphi\partial^a\varphi+\mathcal{R}/2-\mathcal{K}^2/4-\epsilon\right\}\label{eq:2DDGAct}.
\end{align}}
At this point we should comment on how conveniently and surprisingly the above action for a quantum particle, confined to a two-dimensional curved surface, is analogously described by a two-dimensional dilaton gravity, which traditionally arises when describing horizon degrees of freedom of four dimensional black holes. For comparison, the two-dimensional theory that arises in the near horizon of a charged spherically symmetric (Reissner-N\"ordstrom) black hole is given by:
\begin{align}
\begin{split}
&S_{2D}^{RN}=\frac{\ell^2}{4G}\int d^2x\sqrt{-g}e^{-2\varphi}\left\{2\partial_a\varphi\partial^a\varphi+\mathcal{R}-GF^2+\frac{2e^{2\varphi}}{\ell^2}\right\}\label{eq:2DRNS},
\end{split}
\end{align}
where $G$ is Newton's gravitational constant, $\ell$ is the two-dimensional $AdS$ radius and $F$ is the two-dimensionally reduced electromagnetic curvature two-form. 

When comparing the above two actions, \eqref{eq:2DDGAct} and \eqref{eq:2DRNS}, an immediate duality is evident which we summarize in the `da Costa/2D dilaton Gravity' dictionary of Table~\ref{tab:QCP2DDG}.
\begin{table}[htbp!]
\caption{\label{tab:QCP2DDG}Dictionary of da Costa's 2D quantum confined objects (quantities) and their corresponding duals in 2D dilaton gravity.}
\centering 
\resizebox{15.5cm}{!}{
\begin{tabular}{|ll||ll|}
\hline
\multicolumn{2}{|c||}{\textbf{da Costa's 2D Curved Confined Quantum Particle}} & \multicolumn{2}{c|}{\hspace{0.3in}\textbf{Two-Dimensional Near-Horizon Black Hole Dilaton Gravity }}\\\hline\hline
&&&\\
$\ell$ &two-dimensional $AdS$ radius & $\hbar$ &Planck's constant\\
$m$ &confined particle mass& $G$ &Newton's Gravitational Constant\\
$\varphi$ &logarithm of the quantum wave function& $\varphi$ & gravitational dilaton scale of the $S^2$ factor\\
$h_{ab}$ $(h)$ &Euclidian 2D curved embedded metric& $g_{ab}$ $(-g)$ & $s$-wave near-horizon dimensionally reduced black hole\\
$\mathcal{R}$ &semiconductor surface Ricci scalar curvature& $\mathcal{R}$ &near-horizon 2D spacetime Ricci scalar curvature\\
$\mathcal{K}^2/4$ &modulo semiconductor surface intrinsic curvature squared& $GF^2$ &modulo internal Maxwell tensor squared\\
$\epsilon$ &unitless energy eigenvalue& $\frac{2e^{2\varphi}}{\ell^2}$ &cosmological term\\
&&&\\\hline
\end{tabular}}
\end{table}
Next, given the above duality we proceed to calculate the geometrodynamical Einstein equation of \eqref{eq:2DDGAct} via variation with respect to the two-dimensionally curved inverse metric $h^{ab}$. This derivation of the Einstein equation is in general non-trivial, mainly due to the overall $e^{-2\varphi}$ dilatonic weight in the action. This weight in conjunction with Hamilton's principle does not allow us to drop Ricci tensor variations as boundary terms, as one would in the Einstein-Hilbert action principle of four dimensional general relativity {\cite{RRBook,poisson}}. Instead, terms of the order $\delta R_{ab}$ contribute to the final Euler-Lagrange equations of motion. Specifically, our Einstein equation of motion relies on the following properties of the variation of the Riemann tensor and Levi-Civita connection in any dimension:
\begin{align}
\begin{split}\label{eq:RaGVP}
\delta R\indices{^a_b_m_n}=&\nabla_m\delta\Gamma\indices{^a_b_n}-\nabla_n\delta\Gamma\indices{^a_b_m},\\
\delta\Gamma\indices{^a_m_n}=&\frac12\left(-h_{dn}\nabla_m\delta h^{ad}-h_{dm}\nabla_n\delta h^{ad}+h_{sm}h_{dn}\nabla^a\delta h^{sd}\right).
\end{split}
\end{align}
Additionally we note that the Einstein tensor:
\begin{align}
\begin{split}
R_{ab}-\frac12g_{ab}\mathcal{R}=0\label{eg:ETin2D},
\end{split}
\end{align}
is identically zero for any Riemann surface in two dimensions, thus all embedded $h_{ab}$'s {are Einstein metrics} and conformally flat {\cite{nak}}. This implies the following after variation of \eqref{eq:2DDGAct} with respect to $h^{ab}$: 
\begin{align}\label{eq:actvar2d}
\delta S=\frac{\hbar^2}{4m}\int d^2x\sqrt{h}e^{-2\varphi}\left[\left\{-\frac12h_{ab}\left(\partial_c\varphi\partial^c\varphi-\frac{\mathcal{K}^2}{4}-\epsilon\right)+\partial_a\varphi\partial_b\varphi\right\}\delta h^{ab}+\frac12h^{ab}\delta R_{ab}\right].
\end{align}
For the following, let us define:
\begin{align}\label{eq:Tdef}
\delta T=\frac{\hbar^2}{8m}\int d^2x\sqrt{h}\left(e^{-2\varphi}\right)h^{mn}\delta R_{mn},
\end{align}
and using \eqr{eq:RaGVP}, the above becomes:
\begin{align}
\delta T=\frac{\hbar^2}{8m}\int d^2x\sqrt{h}e^{-2\varphi}h^{mn}\left\{\nabla_a\delta\Gamma\indices{^a_m_n}-\nabla_n\delta\Gamma\indices{^a_a_m}\right\},
\end{align}
which we further split, for bookkeeping reasons, into two terms $\delta T=\delta T_1+\delta T_2$, such that:
\begin{align}
\begin{split}
&\begin{cases}
\delta T_1=\frac{\hbar^2}{8m}\int d^2x\sqrt{h}e^{-2\varphi}h^{mn}\nabla_a\delta\Gamma\indices{^a_m_n}\\
\delta T_2=\frac{-\hbar^2}{8m}\int d^2x\sqrt{h}e^{-2\varphi}h^{mn}\nabla_n\delta\Gamma\indices{^a_a_m}
\end{cases}\\
=&\begin{cases}
\delta T_1=\frac{-\hbar^2}{8m}\int d^2x\sqrt{h}\nabla_ae^{-2\varphi}h^{mn}\frac12\left(-h_{dn}\nabla_m\delta h^{ad}-h_{dm}\nabla_n\delta h^{ad}+h_{sm}h_{dn}\nabla^a\delta h^{sd}\right)\\
\delta T_2=\frac{\hbar^2}{8m}\int d^2x\sqrt{h}\nabla_ne^{-2\varphi}h^{mn}\frac12\left(-h_{dm}\nabla_a\delta h^{ad}-h_{da}\nabla_m\delta h^{ad}+h_{dm}\nabla_s\delta h^{sd}\right)
\end{cases},
\end{split}
\end{align}
after reapplication of \eqr{eq:RaGVP} and integration by parts. After some further reindexing, simplifications and another integration by parts, the above becomes:
\begin{align}
\begin{cases}
\delta T_1=\frac{-\hbar^2}{16m}\int d^2x\sqrt{h}\left(2\nabla_d\nabla_ae^{-2\varphi}\delta h^{ad}-\square e^{-2\varphi}h_{dn}\delta h^{dn}\right)\\
\delta T_2=\frac{\hbar^2}{16m}\int d^2x\sqrt{h}\square e^{-2\varphi}h_{ad}\delta h^{ad}
\end{cases},
\end{align}
and thus $\delta T=\delta T_1+\delta T_2$ becomes:
\begin{align}
\delta T=\frac{\hbar^2}{8m}\int d^2x\sqrt{h}\left\{\square e^{-2\varphi}h_{ab}-\nabla_a\nabla_be^{-2\varphi}\right\}\delta h^{ab}.
\end{align}
Now, combining the above result with \eqr{eq:actvar2d} and \eqr{eq:Tdef} and setting $\delta S=0$ we obtain the final version of the full geometrodynamical Einstein field equation:
\begin{align}
\begin{split}
&\left\{-\frac12h_{ab}\left(\partial_c\varphi\partial^c\varphi-\frac{\mathcal{K}^2}{4}-\epsilon\right)+\partial_a\varphi\partial_b\varphi-\frac12\mathcal{K}k_{ab}\right\}e^{-2\varphi}\\
&+\frac12\left(h_{ab}\square e^{-2\varphi}-\nabla_{a}\nabla_{b}e^{-2\varphi}\right)=0\label{eq:2dEEqUT}.
\end{split}
\end{align}
Since, as mentioned earlier, $h_{ab}$ must be conformally flat, the above can be simplified without loss of generality in two dimensions by tracing and thus yielding one single geometrodynamical Einstein equation, given by:
\begin{align}
-\mathcal{K}^2/4+\epsilon+2\nabla_a\varphi\nabla^a\varphi-\square\varphi=0\label{eq:2dEEq}.
\end{align}
This result \eqref{eq:2dEEq} in conjunction with \eqref{eq:DC2d} provides a full quantum mechanical and geometrodynamical description of the da Costa's quantum confined particle:
\begin{align}\label{eq:2dEEqDC2d}
\begin{cases}
-\frac{1}{\sqrt{h}}\partial_a\left(\sqrt{h}h^{ab}\partial_b\psi_{2D}\right)-\left\{\left(\mathcal{K}/2\right)^2-\mathcal{R}/2\right\}\psi_{2D}=\epsilon\psi_{2D}\\
-\mathcal{K}^2/4+\epsilon+2\nabla_a\varphi\nabla^a\varphi-\square\varphi=0
\end{cases}.
\end{align}

{We should briefly comment on the assumption that $\psi_{2D}$ was made to be real in the derivation of \eqref{eq:2dEEq}. A careful and additional very lengthy calculation of the Einstein equation implementing a complex wave function, $\psi=e^{-\varphi}$ and $\psi^*=e^{-\varphi^*}$, yields the following interesting result: 
	\begin{align}\notag
	\begin{split}
	-\frac{\mathcal{K}^2}{4}+\epsilon+\frac12\left(\nabla_a\varphi\nabla^a\varphi+2\nabla_a\varphi\nabla^a\varphi^*+\nabla_a\varphi^*\nabla^a\varphi^*\right)-\frac12\square\left(\varphi+\varphi^*\right)=0,
	\end{split}
	\end{align}
which simplifies to:
	\begin{align}\notag
	\begin{split}
	-\frac{\mathcal{K}^2}{4}+\epsilon+2\nabla_a\left(\frac{\varphi+\varphi^*}{2}\right)\nabla^a\left(\frac{\varphi+\varphi^*}{2}\right)-\square\left(\frac{\varphi+\varphi^*}{2}\right)=0.
	\end{split}
	\end{align}
The above result reproduces \eqref{eq:2dEEq}, {by replacing $\frac{\varphi+\varphi^*}{2}$ with just the real part of $\varphi$}, and implies that the physical (and potentially future measurable) geometrodynamical shape evolution only depends on the real part of the dilaton. This result also implies that we are not limited to considering just real $\psi_{2D}$ in \eqref{eq:2dEEqDC2d}. {Finally, we do not consider geometrodynamical evolution to be equivalent to an adiabatic expansion of the initial geometrical ansatz. Therefore, the resulting new wave functions, when pulled back onto the perturbed curved 2D surfaces, are not necessarily related to the unperturbed ones by a phase factor; instead, they correspond to solutions of different 2D da Costa Schrödinger equations. However, {considering an adiabatic geometric evolution} remains a possibility worth considering, as it could provide valuable insights for future investigations into the quantum geometric analysis of the wave function phase space. That said, we should note that the quantum geometric tensor is an independent quantity from our} $h_{ab}$, which is the physical metric of the 2D curved nano surface.}
\section{Geometric Optimization: A Single Iterative Perturbative Solution}
As a proof-of-concept and for the sake {of obtaining analytic results}, our initial approach here will follow a first iterative perturbative outline given by:
\begin{itemize}
\item First, we select a specific two-dimensional curved surface. Surfaces that are of interest in this area of study include the two-sphere and the two torus {\cite{DASILVA201713,PhysRevLett.100.230403,SCHMIDT2019200}} among others. We will chose the two-sphere for analytical reasons.
\item We then proceed to solve \eqref{eq:DC2d} thereby obtaining an initial eigensystem, i.e. $\left(\varphi_n^0,\epsilon_n^0\right)$. As aforementioned for the two-sphere, the initial eigensystem is obtained analytically and given by spherical harmonics. 
\item Next, we introduce a perturbation to the initially chosen geometry. For example, choosing a two-sphere with metric given by:
\begin{align}
h_{ab}=\left(\partial_a\vec r\right)\cdot\left(\partial_b\vec r\right)=
\left(\begin{array}{cc}1 & 0 \\0 & \sin^2\theta\end{array}\right)
\end{align}
a simple and general perturbation, to $\mathcal{O}(\lambda^3)$, in the radial direction is given by:
\begin{align}
\begin{split}\label{eq:metpert}
h_{ab}^{pert}=
{ \left(\begin{array}{cc}\left(\alpha+\lambda F\right)^2+\lambda^2\left(\partial_\theta F\right)^2 & \lambda^2\left(\partial_\theta F\right)\left(\partial_\phi F\right) \\\lambda^2\left(\partial_\theta F\right)\left(\partial_\phi F\right) & \left(\alpha+\lambda F\right)^2\sin^2\theta+\lambda^2\left(\partial_\phi F\right)^2\end{array}\right)}.
\end{split}
\end{align}
for some small perturbation parameter $\lambda$, arbitrary scale parameter $\alpha$ and unknown function $F(\theta,\phi)$.
\item The next step is to take $\left(\varphi_n^0,\epsilon_n^0\right)$ and $h_{ab}^{pert}$ and insert them into the traced Einstein equation \eqref{eq:2dEEq} for a given $n$ value, thereby obtaining a partial differential equation for $F(\theta,\phi)$ and searching for an analytical solution for the given perturbation. 
\item We then end by demonstrating a plot of this first iterative perturbative shape optimization of $\mathcal{M}^2$, due to a quantum confined particle.
\end{itemize}
In addition to the above outline and flow chart, a full simultaneous numerical solution via the finite element analysis of \eqref{eq:DC2d} and \eqref{eq:2dEEq} is currently under investigation by our research collaboration for forthcoming work, to provide numerical geometric optimization for $\mathcal{M}^2$ due to quantum confinement.  
\subsection{The $s$-wave Case}\label{sec:s-wave}
Choosing the two-sphere as our initial geometry for $\mathcal{M}^2$ and setting $\epsilon=-l(l+1)$ in \eqref{eq:2DSchAct}, results in spherical harmonics, $Y_l^m(\theta,\phi)$ where $n\to (l,m)$, as our initial eigensystem. Substituting the above into \eqref{eq:2dEEq} and solving in the $s$-wave framework ($l=m=0$) and assuming only $\theta$ dependence in the perturbation function $F$ in \eqref{eq:metpert}, yields the following analytical solution to $\mathcal{O}(\lambda^2)$:
\begin{align}
\begin{split}\label{eq:swFf}
&F(\theta)=\alpha\left[\frac{1}{2}+\beta\cos{\theta}+\gamma\left(-1+\cos{\theta}\left(-\frac12 \ln{\left(1-\cos{\theta}\right)}+\frac12\ln{\left(1+\cos\theta\right)}\right)\right)\right],
\end{split}
\end{align}
where $\beta$ and $\gamma$ are integration constants. The integration constants should be dependent upon semiconductor material properties, such as modulus, stress, strain, conductive properties, etc. We have not included those type of contributions in our theoretical framework as yet, and thus, we will choose values for $\alpha$, $\beta$ and $\gamma$ that provide finite solutions to first iteration with respect to the flow chart in Fig.~\ref{fig:FlowChart}, that are scalable to a perturbed unit sphere for proof of concept. 

For the $s$-wave case, we have identified three finite interesting cases depicted in Fig.~\ref{fig:SWp1}-\ref{fig:SWm2} below. For each figure, we implemented the perturbation solution \eqref{eq:swFf} and plotted the resulting perturbed two-sphere. The three cases include values for $\alpha$, $\beta$ and $\gamma$ that range between 0 and 1. The specific choice of $\alpha$ is determined in order to compare our plots to that of a unit two-sphere for visualization purposes. The first case in Fig.~\ref{fig:SWp1}, $\beta$ and $\gamma$ are positive and equal; for the second case in Fig.~\ref{fig:SWm1}, $\beta$ and $\gamma$ are of opposing sign and equal in magnitude; and in the third case, Fig.~\ref{fig:SWm2}, the values for $\beta$ and $\gamma$ differ by a factor of 10.
\begin{figure}[h!] 
	\centering
	\begin{subfigure}{0.2\textwidth}
	\includegraphics[width=1\textwidth]{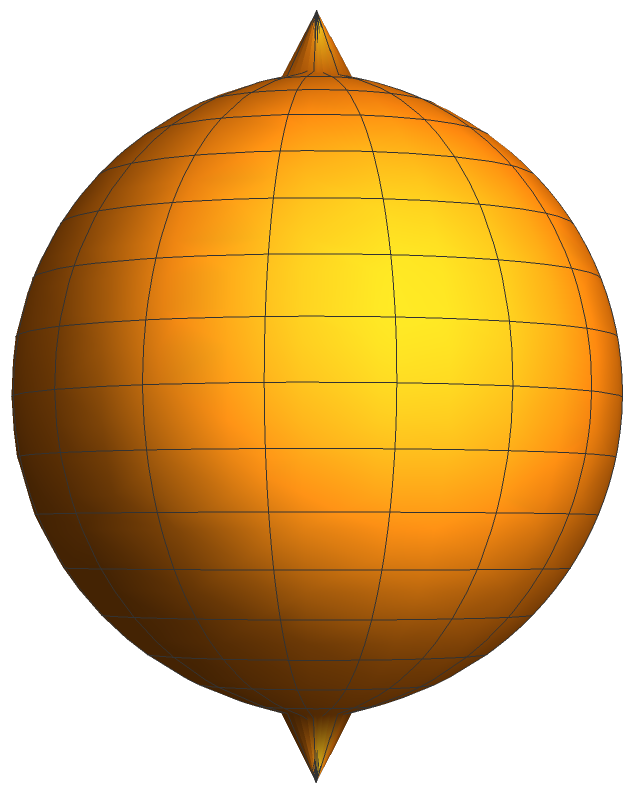}
	\caption{}\label{fig:SWp1a}
	\end{subfigure}~\hspace{1.5cm}~%
	\begin{subfigure}{0.2\textwidth}
    \includegraphics[width=1\textwidth]{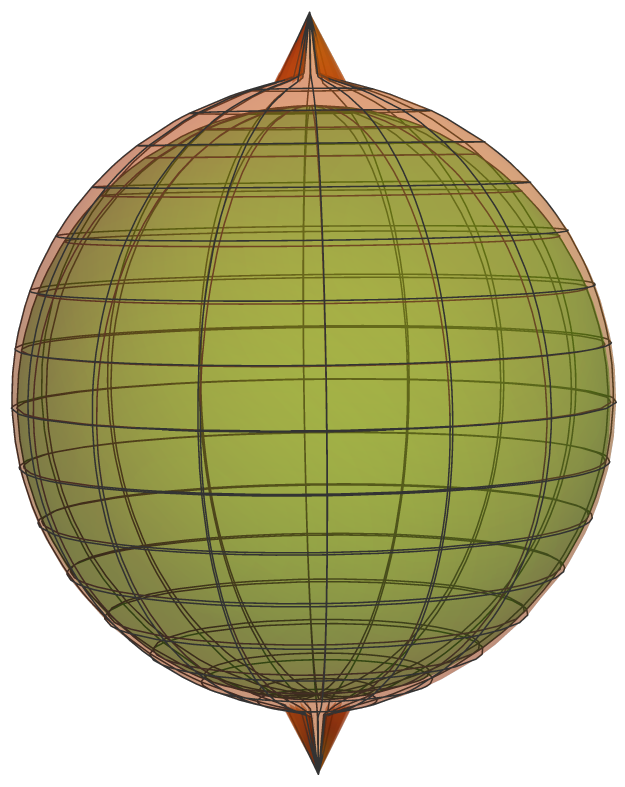}
    \caption{}\label{fig:SWp1b}
	\end{subfigure}
\caption{\label{fig:SWp1}Perturbed geometry \ref{fig:SWp1a} due to a quantum confined particle \emph{\`a la} da Costa, and comparison to initial unperturbed geometry (green) \ref{fig:SWp1b}. The values for $\beta$ and $\gamma$ are positive, equal and bounded between 0 and 1.}
  \end{figure}
\begin{figure}[h!] 
	\centering
	\begin{subfigure}{0.2\textwidth}
	\includegraphics[width=1\textwidth]{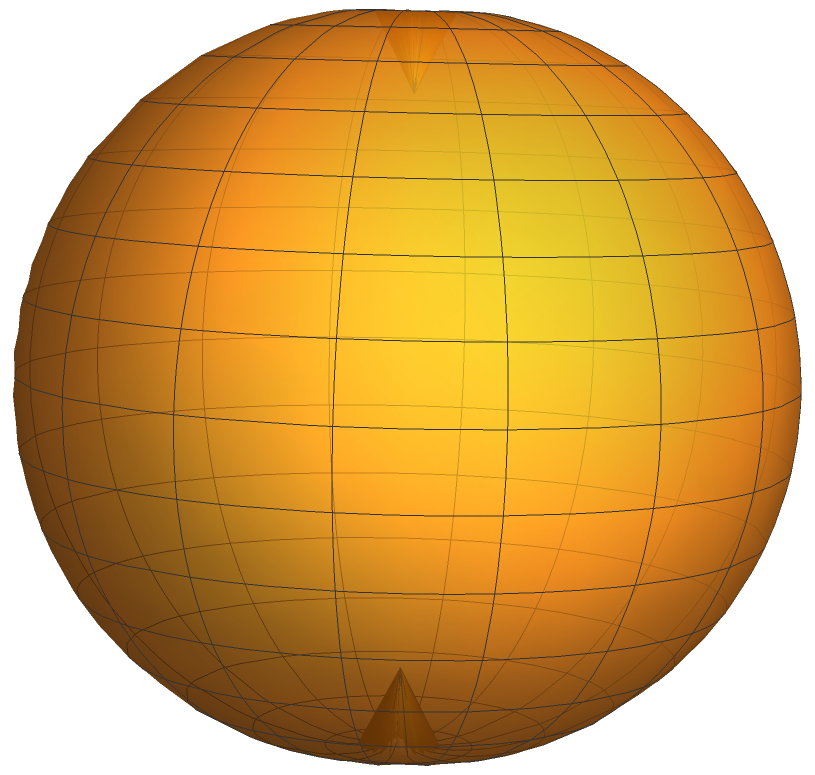}
	\caption{}\label{fig:SWm1a}
	\end{subfigure}~\hspace{.75cm}~%
	\begin{subfigure}{0.2\textwidth}
    \includegraphics[width=1\textwidth]{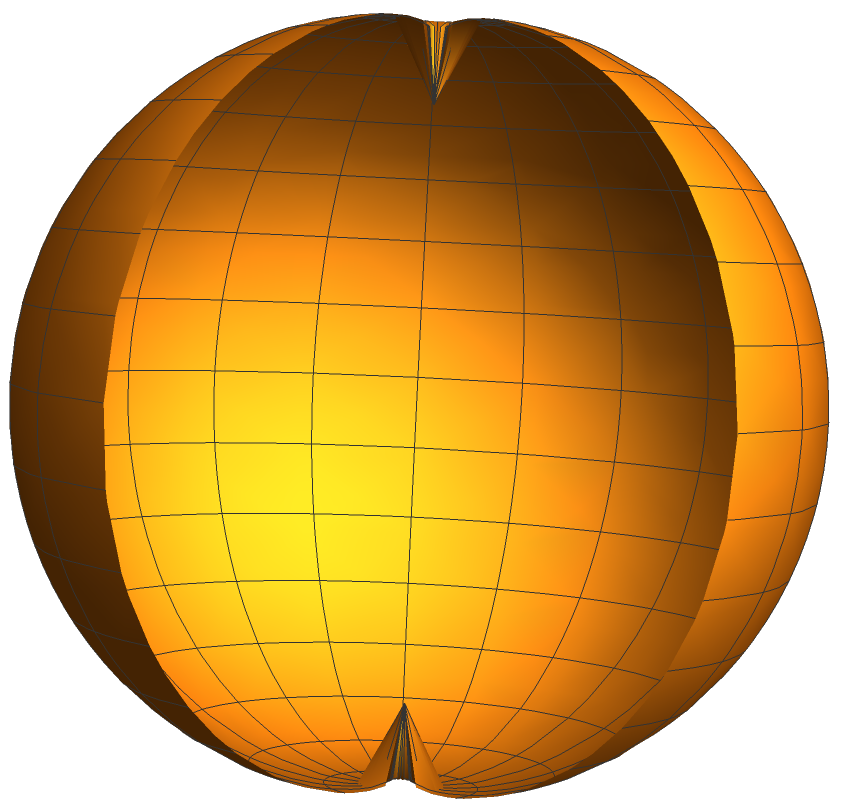}
    \caption{}\label{fig:SWm1b}
	\end{subfigure}~\hspace{.75cm}~%
	\begin{subfigure}{0.2\textwidth}
    \includegraphics[width=1\textwidth]{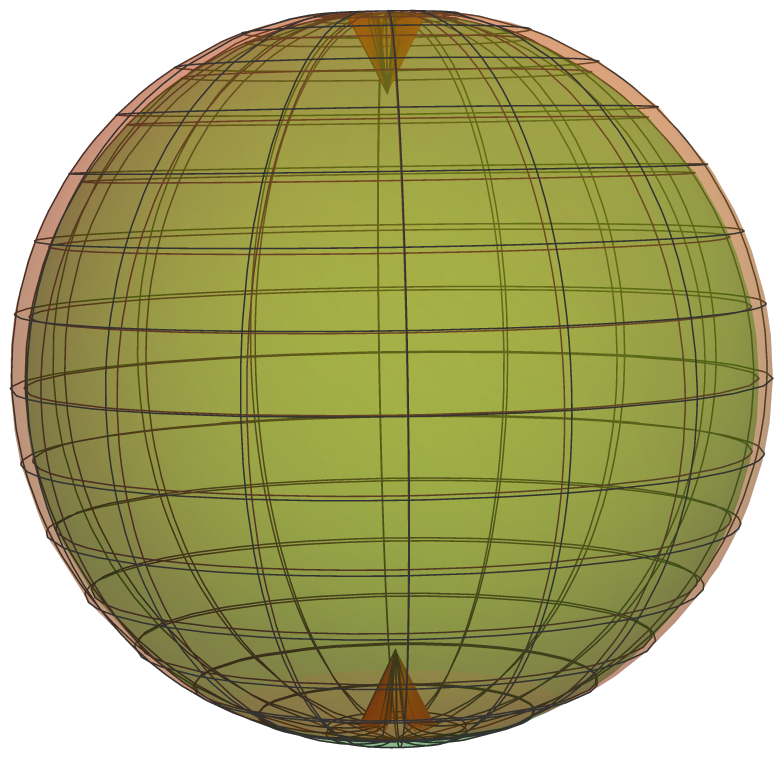}
    \caption{}\label{fig:SWm1c}
	\end{subfigure}
\caption{\label{fig:SWm1}Perturbed geometry \ref{fig:SWm1a} and its cross sectional plot \ref{fig:SWm1b} (displaying the interior) due to a quantum confined particle \emph{\`a la} da Costa, and comparison to initial unperturbed geometry (green) \ref{fig:SWm1c}. The values for $\beta$ and $\gamma$ are opposite in sign, equal in magnitude and bounded between 0 and 1.}
  \end{figure}
\begin{figure}[h!] 
	\centering
	\begin{subfigure}{0.2\textwidth}
	\includegraphics[width=1\textwidth]{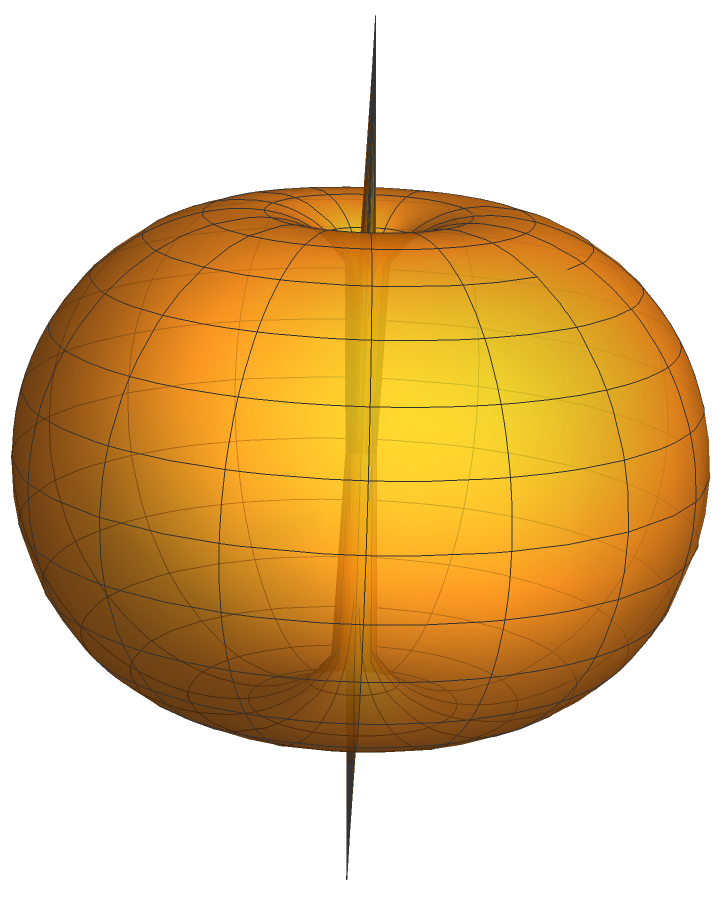}
	\caption{}\label{fig:SWm2a}
	\end{subfigure}~\hspace{.75cm}~%
	\begin{subfigure}{0.2\textwidth}
    \includegraphics[width=1\textwidth]{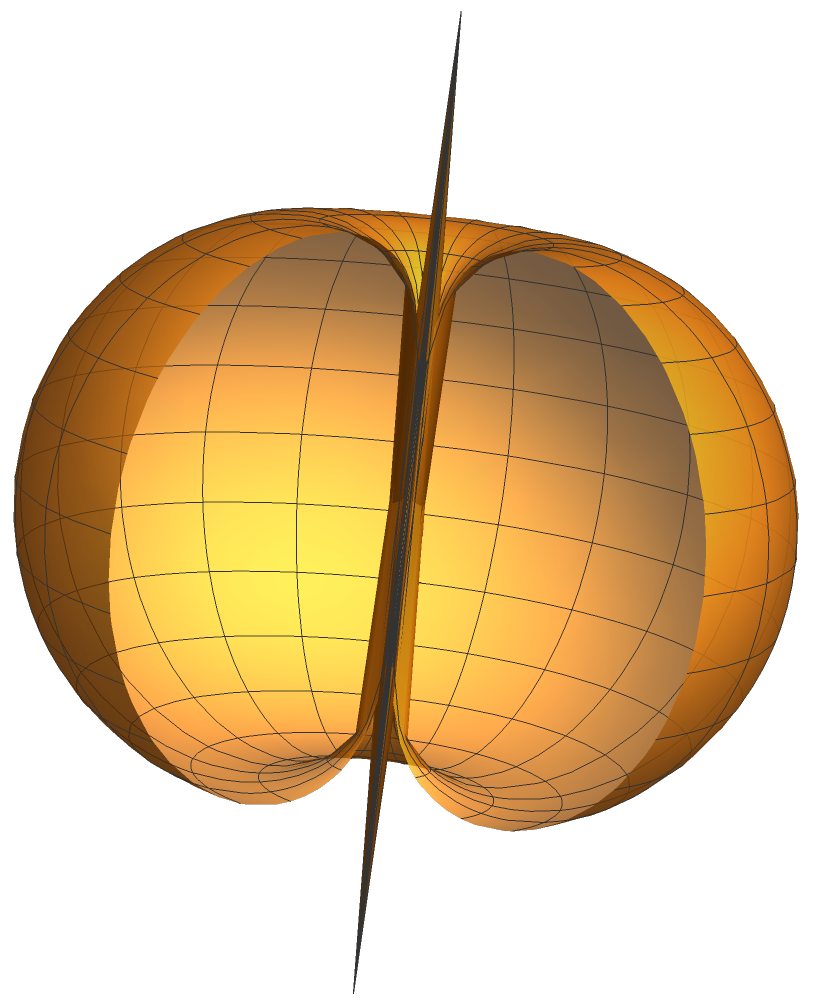}
    \caption{}\label{fig:SWm2b}
	\end{subfigure}~\hspace{.75cm}~%
	\begin{subfigure}{0.2\textwidth}
    \includegraphics[width=1\textwidth]{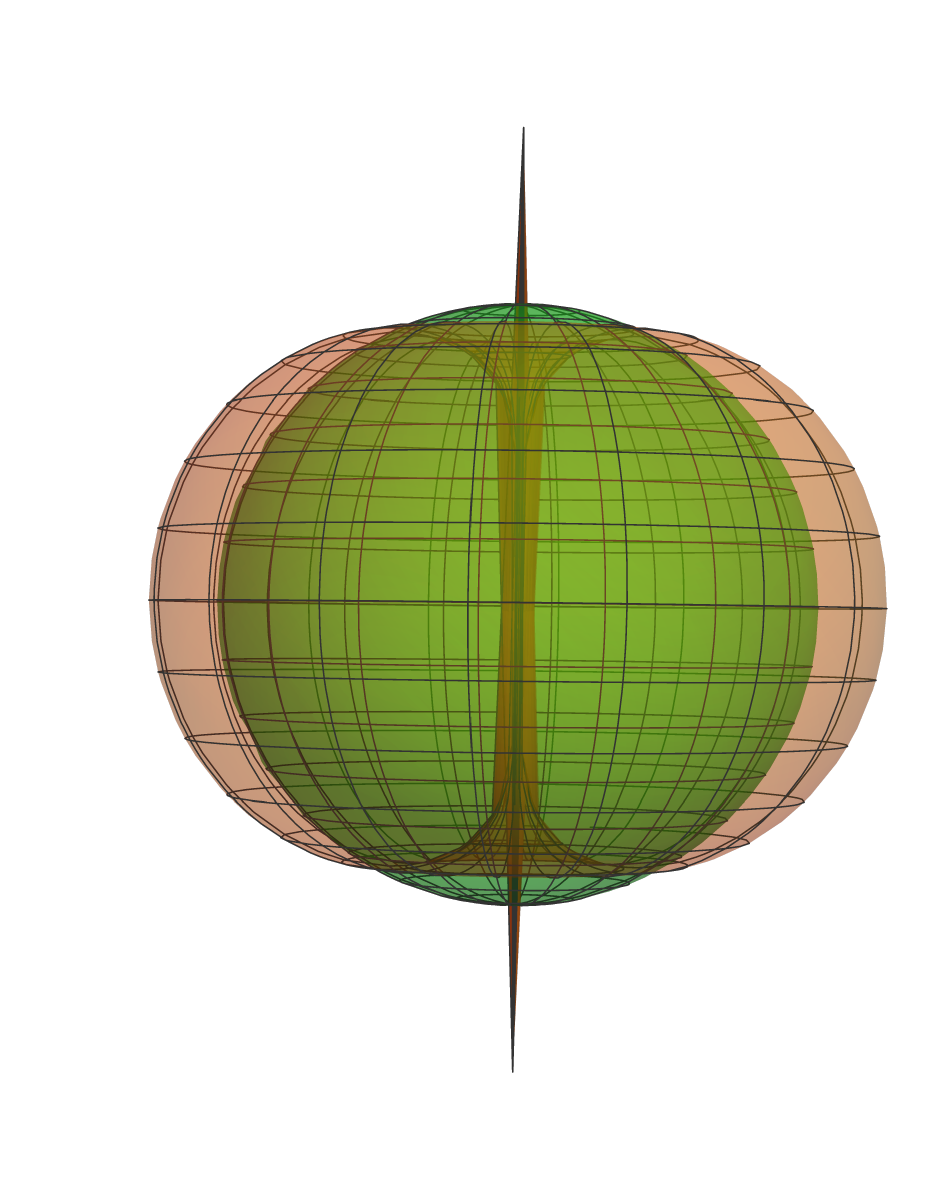}
    \caption{}\label{fig:SWm2c}
	\end{subfigure}
\caption{\label{fig:SWm2}Perturbed geometry \ref{fig:SWm2a} and its cross sectional plot \ref{fig:SWm2b} (displaying the interior) due to a quantum confined particle \emph{\`a la} da Costa, and comparison to initial unperturbed geometry (green) \ref{fig:SWm2c}. The values for $\beta$ and $\gamma$ are opposite in sign, differ by a factor of 10 in magnitude and bounded between 0 and 1.}
  \end{figure}

From Fig.~\ref{fig:SWp1}-\ref{fig:SWm2} it is apparent that there is a geometrodynamical feedback from the quantum confined particle's eigensystem onto the initially chosen geometry of $\mathcal{M}^2$. 
\subsection{The $l=1$ and $m=\pm1$ Cases}
\begin{figure}[h!] 
	\centering
	\begin{subfigure}{0.2\textwidth}
	\includegraphics[width=1\textwidth]{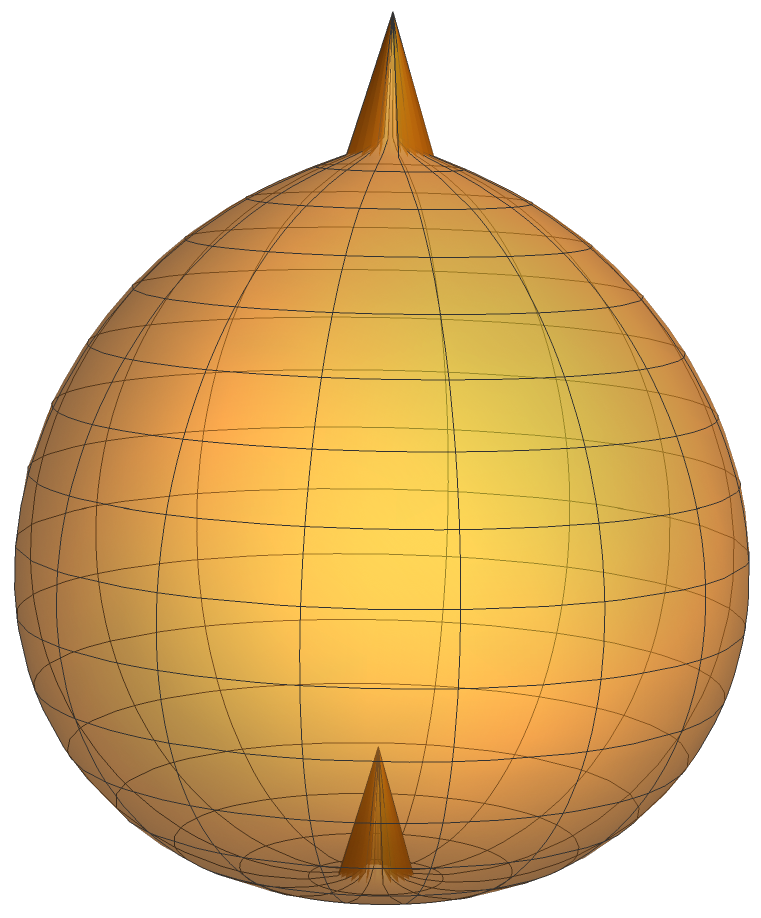}
	\caption{}\label{fig:l1m1p1a}
	\end{subfigure}~\hspace{1.5cm}~%
	\begin{subfigure}{0.2\textwidth}
    \includegraphics[width=1\textwidth]{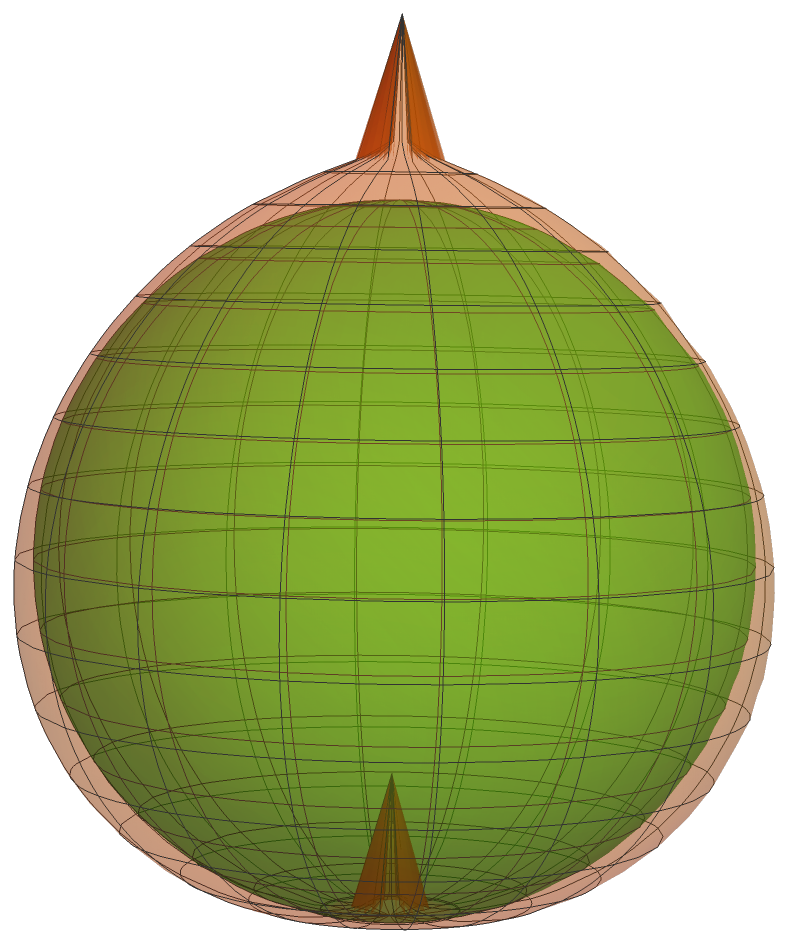}
    \caption{}\label{fig:l1m1p1b}
	\end{subfigure}
\caption{\label{fig:l1m1p1}Perturbed geometry \ref{fig:l1m1p1a} due to a quantum confined particle \emph{\`a la} da Costa, and comparison to initial unperturbed (green) geometry \ref{fig:l1m1p1b}. The values for $\beta$ and $\gamma$ are positive, equal and bounded between 0 and 1.}
  \end{figure}

\begin{figure}[h!] 
	\centering
	\begin{subfigure}{0.2\textwidth}
	\includegraphics[width=1\textwidth]{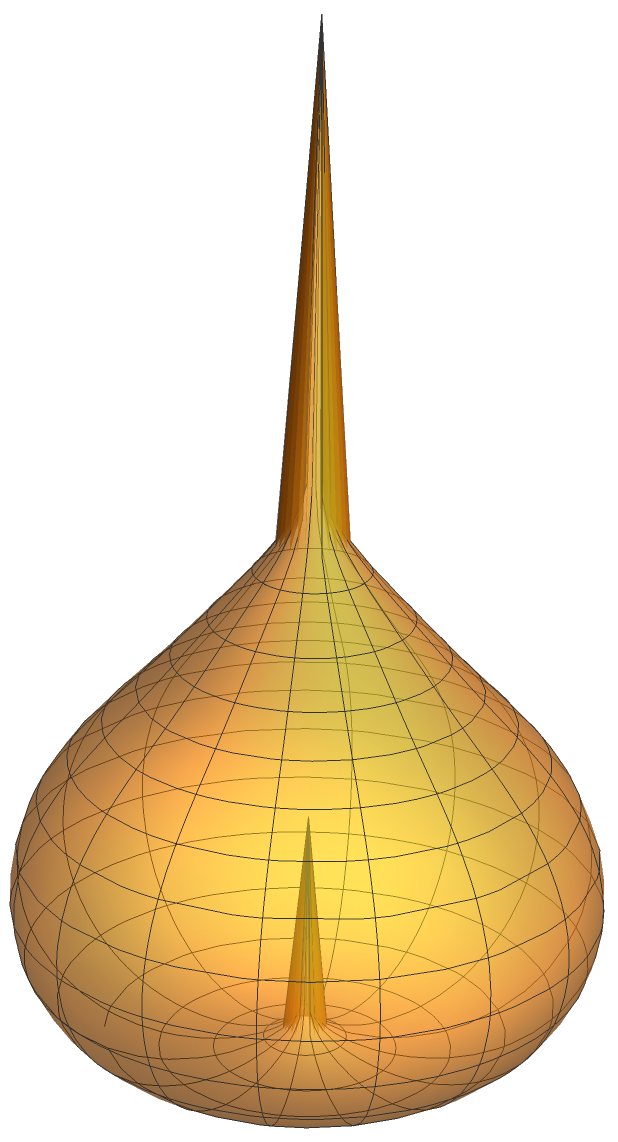}
	\caption{}\label{fig:l1m1p2a}
	\end{subfigure}~\hspace{1.5cm}~%
	\begin{subfigure}{0.2\textwidth}
    \includegraphics[width=1\textwidth]{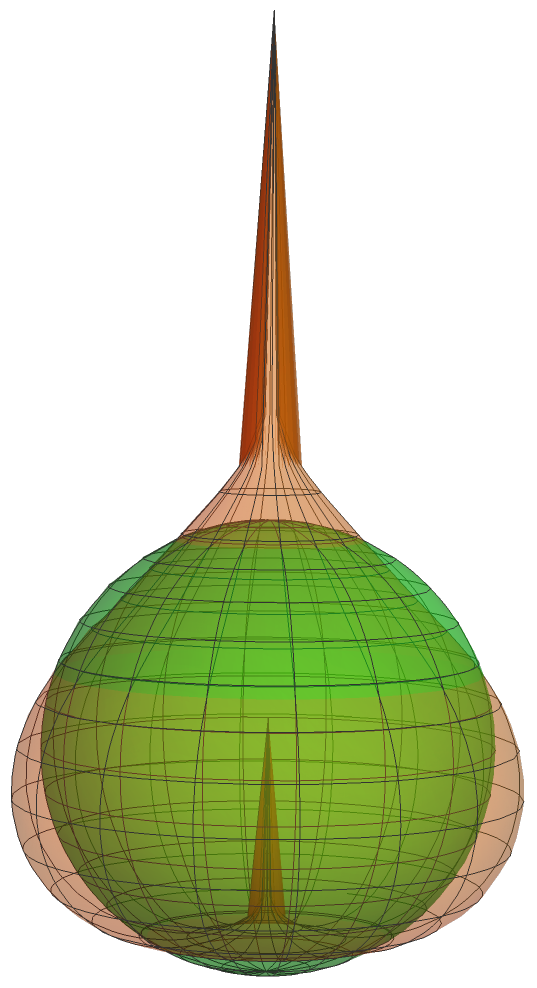}
    \caption{}\label{fig:l1m1p2b}
	\end{subfigure}
\caption{\label{fig:l1m1p2}Perturbed geometry \ref{fig:l1m1p2a} due to a quantum confined particle \emph{\`a la} da Costa, and comparison to initial unperturbed geometry \ref{fig:l1m1p2b}. The values for $\beta$ and $\gamma$ are positive, differ by a factor of 10 and bounded between 0 and 1.}
  \end{figure}

Two of the $l=1$ cases are another example where we are able to obtain analytical solutions for the perturbation function $F$, assuming only $\theta$ dependence and following the same procedures from the previous subsection~\ref{sec:s-wave}. For the $m=0$ case, we were not able to find a closed form analytical solution, but the other two cases $m=\pm1$, there exists one common and unique analytical closed form solution in terms of Legendre polynomials ($P_n(x)$) and Legendre functions of the second kind ($Q_n(x)$), given by:
\begin{align}
\begin{split}\label{eq:11Ff}
F(\theta)=&\alpha\left[2+\alpha^2+\beta P_{(-1+\sqrt{33})/2}(\cos\theta)+\gamma Q_{(-1+\sqrt{33})/2}(\cos\theta)\right].
\end{split}
\end{align}
For this specific case, we display two sets of plots for choices of the scale parameter $\alpha$ and integration constants $\beta$ and $\gamma$. In the first plot, displayed in Fig.~\ref{fig:l1m1p1}, the integration constants are equal and between 0 and 1. Varying the constants in sign only affects the direction of the polar peaks along the $z$-axis. The overall pumpkin shape seems universally unaffected. In Fig.~\ref{fig:l1m1p2} the integration constants differ by a factor of 10, but still sampled between 0 and 1. Both perturbed geometries are compared to a green unit two-sphere in the second plot of each case. 

Although the integration constants still need to be tuned to reflect the actual material properties, {we have shown in a significant number of cases that geometric deformation of the two-dimensional curved surfaces arises} due to the feedback of a quantum confined particle.
\section{External Fields}
Finally and as proposed above, we will detail and outline how to incorporate the presence of external electromagnetic fields in our formalism of geometric shape optimization. The presence of external fields is very important, since they can be used to tune desired outcomes and appear naturally in the design of electronic devices. However, our unique dual description allows us to unveil new phenomena and rich physics. This is mainly due to our action integral approach, motivated from the finite element analysis paradigm, and resulting gravity analog detailed in Table~\ref{tab:QCP2DDG}. To begin, let us start by outlining a simple example of a unit two-sphere centered at the origin with an external constant magnetic field along the $z$-axis, i.e.:
\begin{align}
\begin{split}
h_{ab}=&
\left(\begin{array}{cc}1 & 0 \\0 & \sin^2\theta\end{array}\right),~\text{and}~\vec B=B_0\hat k.
\end{split}
\end{align}
One choice for a three vector potential whose curl yields the above magnetic field is given by the Landau gauge {\cite{LLQM}}:
\begin{align}
\vec A=&-B_0y\hat \imath.
\end{align}
Now, to incorporate this external magnetic field in da Costa's formalism, given by \eqref{eq:DC2d}, we need to first augment the operator $\partial_a$ by the pullback of $\vec A$ given by $A_a=\partial_a\vec r\cdot\vec A$ and thus defining the gauge covariant derivative, in the usual way, on the two-dimensional curved surface: $D_a=\nabla_a-ieA_a$. 

\begin{figure}[htbp!]
\centering
\includegraphics[width=0.475\textwidth]{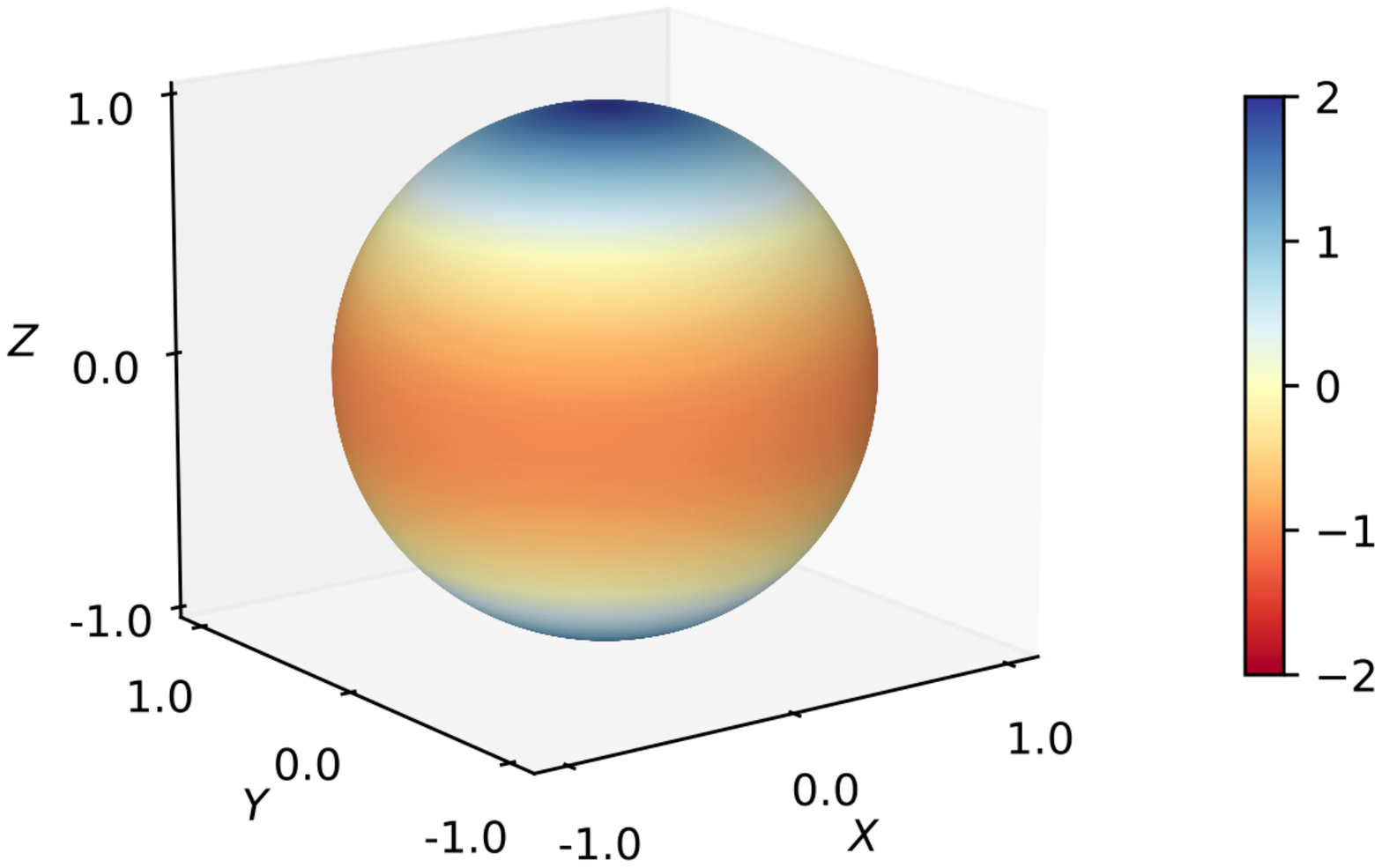}
\caption{Color-coded surface density plot of $J^\phi$ over the two-sphere.}\label{fig:densplot}
\end{figure}

This is standard practice, but has not been addressed before in the context of pulling down the three vector $\vec A$ to the two-dimensional curved surface. This process also induces a circulating surface magnetic field sourced by antiparallel surface currents, which are projected and distributed nontrivially on the respective embedded curved surface. Let us demonstrate; first, the pullback of the three vector potential in the above simple example gives rise to the embedded electromagnetic field tensor (curvature two-form) given by:
\begin{align}
\begin{split}
F_{ab}=&\partial_aA_b-\partial_bA_a= \left(\begin{array}{cc}0 &-B_\theta \\B_\theta& 0\end{array}\right)\\
=&\left(\begin{array}{cc}0 &-B_0\cos\theta\sin\theta \\B_0\cos\theta\sin\theta& 0\end{array}\right)
\end{split}
\end{align}
which (keeping in mind that $F_{ab}$ is purely spatial) reveals the induced embedded two-dimensional specially dynamic magnetic field. The source of this induced magnetic field is a current density nontrivially distributed across the curved two-dimensional surface, depicted in Fig.~\ref{fig:densplot}, which can be determined by the Maxwell equation:
\begin{align}
\nabla_bF^{ab}=-J^a,
\end{align}
where $\nabla_a$ is the Levi-Civita covariant derivative and $J^a=\left(J^\theta,J^\phi\right)$ is the conserved $U(1)$ N\"other current density on the curved embedded surface. 

A simple calculation following from the above yields the surface current density:
\begin{align}
J^a=\left(0,B_0\left(2\cos^2\theta-\sin\theta\right)\right),
\end{align}
whose normalized plot is given in Fig~\ref{fig:pbcharge}.
\begin{figure}[htbp!]
\centering
\begin{tikzpicture}
\begin{axis}[domain=0:pi, samples=100]
\addplot [no marks, color=blue, thick]{2*(cos(x*180/pi))^2-sin(x*180/pi)}; 
\end{axis}
\draw[] (-.5,3) node[above,rotate=90]{Induced Current Density, $J^\phi$};
\draw[] (3.5,-.5) node[below]{$\theta$ in radians};
\end{tikzpicture}
\caption{Plot of the induced surface current density $J^\phi$ over the range of $\theta$. We see that the charge density has its maximum at the poles where the external magnetic field pierces the two-sphere, it has a negative value on the equator and is zero at $\sim\pi/3.5066$ and $\sim \pi/2.3275+\pi/3.5066$.}\label{fig:pbcharge}
\end{figure}
We see that in the presence of a constant external magnetic field, the quantum confined particle (to a two-dimensional curved surface) experiences both attractive and repulsive dynamical Lorenz force interactions, which will feed back upon the geometrodynamics of the curved surface. To incorporate these new features, the action principle of \eqref{eq:2DDGAct} needs to be augmented in the following form:
\begin{align}
\begin{split}
S=&\frac{\hbar^2}{2m}\int d^2x\sqrt{h}e^{-(\varphi^*+\varphi)}\left\{D^*_a\varphi^* D^a\varphi+\mathcal{R}/2-\mathcal{K}^2/4-\epsilon-\frac{1}{4}\Phi F^2\right\}\label{eq:2DDGActMAX},
\end{split}
\end{align}
where we now consider a complex dilaton, $\Phi=2me^{(\varphi^*+\varphi)}/\hbar^2$ and $F^2=F_{ab}F^{ab}$. This necessary augmentation will alter the resulting geometrodynamical Einstein equation of motion \eqref{eq:2dEEq}, given the $U(1)$ energy momentum tensor in terms of its functional generator:
\begin{align}
\frac{-2}{\sqrt{h}}\frac{\delta S_{U(1)}}{\delta h^{ab}}\equiv T^{U(1)}_{ab}=-\frac14h_{ab}F^2+h^{cd}F_{ac}F_{bd},
\end{align}
and adds additional new features to the shape optimization procedures outlined in previous sections. Additionally, the variation of the above total action \eqref{eq:2DDGActMAX} with respect to the pullback field $A_a$ will give rise to an additional equation of motion or constraint, given the functional generator equation for $U(1)$ symmetry: 
\begin{align}
\frac{\delta S}{\delta A_a}\equiv J^a.
\end{align}
This is in part due to the gauge covariant derivative $D_a$ exhibiting $A_a$ dependence and where
\begin{align}
\Phi J_a=ie\left(\varphi^*\nabla_a\varphi-\varphi\nabla_a\varphi^*\right)+2e^2\varphi^*A_a\varphi
\end{align}
is the conserved induced surface N\"other current obtained above in terms of the complex dilaton. This new feature, which is also missing in previous research endeavors, will help and allow for more numerical constraints in future shape optimization pursuits. 

To summarize without loss of generality, let us consider the simplified case for a real dilaton $\varphi^*=\varphi$. In this case the total geometrodynamical shape optimization problem reduces to the simultaneous solution of the following equations of motion:
\begin{align}
\begin{cases}\label{eq:FullMaxExEQM}
-\frac{\hbar^2}{2m}\frac{1}{\sqrt{h}}D^*_a\left(\sqrt{h}h^{ab}D_b\psi_{2D}\right)-\frac{\hbar^2}{2m}\left\{\left(\mathcal{K}/2\right)^2-\mathcal{R}/2\right\}\psi_{2D}=E_{2D}\psi_{2D}&\text{{\footnotesize daCSEq}}\\
-\mathcal{K}^2/4+\epsilon+2\nabla_a\varphi\nabla^a\varphi-\square\varphi+\frac{m}{4\hbar^2}F^2=0&\text{{\footnotesize EEq}}\\
\nabla_bF^{ab}=-J^a&\text{{\footnotesize MEq}}
\end{cases},
\end{align}
which are the da Costa-Schr\"odinger equation with induced curvature surface potentials, the Einstein equation with induced $U(1)$ contribution and the pulled-back induced Maxwell equation. We should note here that despite starting this section with a simplistic example, the above shape optimization equations are not specific to any given example of a quantum confined particle within a given external field configuration, but instead it is a recipe for any general curved two-dimensional surface in any arbitrary external field configuration.  
\section{{Future Work: Full Numerical Perturbative Shape Optimization}}\label{sec:FW}
{Given the results from the previous two sections, we are motivated and encouraged to perform a more detailed numerical shape optimization investigation, which is currently underway with results forthcoming. When performing numerical investigations, our main computational tool, for solving derived field equations, is rooted in high accuracy finite element analysis within the framework of the action integral principle \cite{Ram}, which in part motivated the action construction in \eqref{eq:2DSchAct}.} In this method, the device geometry is discretized into refined finite element meshes. Within each element the potential functions are expressed as linear combination of ($n_{int}$) Hermite interpolation polynomials multiplied by as-yet undetermined coefficients. {The action integral can {now be spatially} integrated.} Employing the principle of stationary action, variation of the action with respect to these undetermined coefficients results in a set of linear equations, which are then solved using sparse matrix solvers. For example, take a one dimensional action $S=S(x)$ and a spatial discretization into $n$ elements:
\begin{align}
S=\sum_{i_{el}}^nS^{\left(i_{el}\right)}.
\end{align} 
Additionally, we also discretize and interpolate the wave function $\psi(x)$ via Hermite interpolation polynomials per element, $N_j^{i_{el}}(x)$ such that:
\begin{align}
\psi(x)=\sum_{i_{el},j}\psi^{i_{el}}_jN_j^{i_{el}}(x),
\end{align} 
where $\psi^{i_{el}}_j$ are interpolation constants and all the spatial dependence is now contained in the local interpolation polynomials. Now, consider a simple minimally coupled action with kinetic term given by: 
\begin{align}
S_{K}^{\left(i_{el}\right)}=\sum_{j,k}^{n_{int}}\psi^*_{j}\left[\int dxN'_j(x)N'_k(x)\right]\psi_{k},
\end{align}
per element. Then the integral can be performed over $x$ employing Gauss quadratures. Performing the relevant numerics, we are left with what we call the global action over all  the terms and elements: 
\begin{align}
S=\sum_{\alpha,\beta}^{n_{glob}}\psi^*_{\alpha}\left(\mathcal{M}_{\alpha\beta}-E\mathcal{B}_{\alpha\beta}\right)\psi_{\beta},
\end{align}
with equation of motion given by the generalized eigenvalue problem:
\begin{align}
\delta_{\psi^*_\alpha}S=\sum_{\beta}\left(\mathcal{M}_{\alpha\beta}-E\mathcal{B}_{\alpha\beta}\right)\psi_{\beta}=0,
\end{align}
for the global matrix $\mathcal{M}_{\alpha\beta}$ and overlap matrix $\mathcal{B}_{\alpha\beta}$. In other words, given an action the solution to its field equation becomes an eigenvalue problem within the action level finite element method, this motivates an action principle for the pulled back Schr\"odinger equation in \eqr{eq:DC2d}, which we presented in \eqref{eq:2DSchAct} and its gravitational dual in \eqref{eq:2DDGAct} and finally its dual including external electromagnetic fields in \eqref{eq:2DDGActMAX}.

{We have shown that FEM delivers double-precision accuracy \cite{Ram}. As proposed above, we are interested in using \eqref{eq:FullMaxExEQM} as a powerful tool for the study of shape optimization in this particular physical application}. A full iterative perturbative approach using finite element analysis and including external fields, is outlined in detail in the flowchart depicted in Fig.~\ref{fig:FlowChart}.
\begin{figure}[htbp!]
	\centering\scalebox{.75}{
	\begin{tikzpicture}[scale=.5, node distance=3.cm]

	\node (start) [start,xshift=-10cm] {{\bf Start} by Choosing a 2D Metric Ansatz $h_{ab}$};
	\node (hpert) [start, below of=start,yshift=-0.5cm] {Perturb the Geometry $h^{Pert}_{ab}=h_{ab}+f_{ab}(\lambda, \theta, \phi)$};
	\node (FEM) [start, right of=start, xshift=3cm] {Solve \eqref{eq:DC2d} Using FEM};
	\node (Eigen) [start, right of=hpert, xshift=3cm] {Obtain Initial Unperturbed Eigensystem ($\varphi^0_n, \epsilon^0_n$)};
	\node (EE) [goal1, below of=hpert, xshift=2cm,yshift=-1cm] {{\bf Goal 1:} Solve for Metric Perturbation $f_{ab}(\lambda, \theta, \phi)$ by Feeding $h^{Pert}_{ab}, (\varphi^0_n, \epsilon^0_n)$ into \eqref{eq:2dEEq}};

	\node (pro2a) [goal1, below of=FEM, xshift=-3.95cm,yshift=-8.75cm] {Determine the Full Perturbed Geometry $h^{Pert}_{ab}$ and Treat as a New Ansatz };

	\node (EM) [goal3, right of=EE, xshift=2cm] {{\bf Goal 3:} Add External E\&M Fields};
	\node (Potential) [goal3, below of=EM] {Determine Induced Surface Potential if Goal 3 added};
	\node (stop) [goal2, left of=pro2a, xshift=-2cm,yshift=2cm] {{\bf Goal 2:} Reevaluate until a Converging Optimum Shape};

	\draw [arrow] (start) -- (hpert);
	\draw [arrow] (start) -- (FEM);
	\draw [arrow] (FEM) -- (Eigen);
	\draw [arrow] (Eigen) -- (EE);
	\draw [arrow] (hpert) -- (EE);
	\draw [arrow] (pro2a) -- (Potential);
	\draw [arrow] (EM) -- (EE);
	\draw [arrow] (EE) -- (pro2a);
	\draw [arrow] (pro2a) -- (stop);
	\draw [arrow] (stop) |- (start);

	\end{tikzpicture}}
	\caption{\label{fig:FlowChart}Flowchart of an iterative perturbative shape optimization using \eqref{eq:FullMaxExEQM}, assuming convergence.}
\end{figure}
{As stated in the caption of Fig.~\ref{fig:FlowChart}, we are assuming the proposed full iterative perturbative shape optimization should converge. This is not a naive assumption, but is rooted robustly within the differential geometric nature and solutions space of quantum gravity in two dimensions. As discussed in the introduction and in \eqref{eg:ETin2D}, $h_{ab}$ must be conformally flat, i.e.:}
\begin{align}
h_{ab}=\Omega^{-2}\eta_{ab},
\end{align}
{where $\Omega^{-2}$ is an arbitrary conformal factor mapping $h_{ab}$ to the flat (zero curvature) metric $\eta_{ab}$. In other words, regardless of how many iterative perturbations (according to Fig.~\ref{fig:FlowChart}) are performed, the resulting optimized metric, $h^{opt}_{ab}$, must also be conformally flat:}
\begin{align}
h^{opt}_{ab}=\Omega^{-2}_{opt}\eta_{ab}.
\end{align}
{Thus, the final optimized metric should converge to the initial (zeroth-order) metric modulo a finite conformal factor given by the relationship:}
\begin{align}
h^{opt}_{ab}=\frac{\Omega^{2}}{\Omega^{2}_{opt}}h_{ab}.
\end{align}
\section{Conclusion}
The {important} work and contribution 
of R. C. T. da Costa cannot be overstated in initiating a near revolution before their time. This pioneering work combined the arts of differential geometry and quantum mechanics before the advent of string theory, string theory's $AdS/CFT$ correspondence, $AdS/CMT$ correspondence and the general gauge gravity correspondence. In fact, the majority of research explosion based upon this paradigm has {been} spawned only in the last decade, as discussed in the introduction. 

{The majority of earlier work aims at extending and solving numerically or analytically many different cases of the da Costa Schr\"odinger equation in \eqref{eq:DC2d}, for fixed two-dimensional curved geometries}. {In this article we have formulated the first} novel extension by showing how to implement the da Costa formalism as a basis for a geometrodynamical shape optimization of the two-dimensional curved surfaces with an initial quantum confined eigensystem, a fundamental question and problem in the current semiconductor device design and manufacturing field of research\cite{electricMatRev}.

In this work, we have performed an initial proof-of-concept perturbative calculation of how the two-dimensional curved geometry would be deformed given by {the feedback} exhibited by the initial confined eigensystem. Additionally, we have laid the basic formulation for how to extend our work within a full numerical iterative approach with any given external electromagnetic field configuration. In doing so, we have relied on {the} finite element analysis method, which is rooted within a stationary action principle. 

Some theoretical considerations and comments are due. {As aforementioned,} the action in \eqref{eq:2DSchAct} is not a classical one, but should be interpreted as a quantum effective one, since the variation of \eqref{eq:2DSchAct} with respect to the wave function $\psi$ yields the quantum mechanical Schr\"odinger equation. In other words, \eqref{eq:2DSchAct} is the effective action defined in the canonical way:
\begin{align}
S=\Gamma_{\rm{eff}}(\psi)=-i\ln{(Z)},
\end{align}
where 
\begin{align}
Z=\int \mathcal{D}x e^{iS_{cl}(x)}
\end{align}
is the partition function of the classical action, $S_{cl}(x)$. {Now, this implies that the Einstein equation in \eqref{eq:2dEEqUT} contains a quantum mechanical energy momentum tensor on its left side, given by}:
\begin{align}
\begin{split}
\langle T_{ab}\rangle =&\left\{-\frac12h_{ab}\left(\partial_c\varphi\partial^c\varphi-\frac{\mathcal{K}^2}{4}-\epsilon\right)+\partial_a\varphi\partial_b\varphi+\right.\\
&\left.-\frac12\mathcal{K}k_{ab}\right\}e^{-2\varphi}+\frac12\left(h_{ab}\square e^{-2\varphi}-\nabla_{a}\nabla_{b}e^{-2\varphi}\right)
\end{split},
\end{align}
since 
\begin{align}
\begin{split}
\langle T_{ab}\rangle =&\int \mathcal{D}x \left(T_{ab}\right) e^{iS_{cl}(x)}=\int \mathcal{D}x \left(\frac{-2}{\sqrt{h}}\frac{\delta}{\delta h^{ab}}\right) e^{iS_{cl}(x)}\\
=&\frac{-2}{\sqrt{h}}\frac{\delta}{\delta h^{ab}}\int \mathcal{D}x e^{iS_{cl}(x)}=\frac{-2}{\sqrt{h}}\frac{\delta}{\delta h^{ab}}Z\\
=&iZ\frac{-2}{\sqrt{h}}\frac{\delta}{\delta h^{ab}}\Gamma_{\rm{eff}}(\psi),
\end{split}
\end{align}
where we have used the functional generator of $T_{ab}$ and the wave function $\psi$ acts like an auxiliary field for the sake of locality restoration. The above yields the equation of motion in \eqref{eq:2dEEqUT} modulo an arbitrary normalization of $iZ$. Additionally, $\langle T_{ab}\rangle$ above is a renormalized energy momentum tensor (EMT), since it satisfies a Bianchi identity.  However, an important question for the semiconductor and materials science researchers is, whether a combination of the above renormalized stress tensor is conserved or not. In other words, what combination of the above EMT contributes to the two-dimensional gravitation anomaly cancelation. Dissecting this detail in \eqref{eq:2dEEqUT} would be a first step towards isolation and incorporation {of} material properties of the two-dimensional curved surface into the shape optimization procedure. This is a matter for future research and work. 
\ack{SR and LR would like to thank the Grinnell College Harris Fellowship Foundation for supporting this work and also WPI for their hospitality where this work was initially conceived. LR would like to thank Vincent Rodgers for
support, encouragement and very enlightening discussions. LRR thanks PK Aravind for enlightening discussions.

We also thank our referees for their appreciated work, comments and suggestions, which greatly improved the paper's presentation, accuracy and readability.  
\appendix
\section*{References}
\bibliography{cftgr}
\bibliographystyle{utphys}
\end{document}